\documentclass[aps,prl,lengthcheck,superscriptaddress,showpacs]{revtex4-1}
\usepackage{graphicx}
\usepackage{amsmath}

\begin{document}

\title{Dramatic Enhancement of Third-Harmonic Generation in Plasmonic Nanostructures via Nonlocal Effects}

\author{Cristian Cirac\`i}
\email[]{cristian.ciraci@duke.edu}

\affiliation{Center for Metamaterials and Integrated Plasmonics, and Departement of Electrical and Computer Engineering, Duke University, Durham NC 27708, U.S.}

\author{Michael Scalora}
\affiliation{C.M. Bowden Research Facility, US Army, RDECOM, Redstone Arsenal, AL 35803, U.S.}

\author{David R. Smith}
\affiliation{Center for Metamaterials and Integrated Plasmonics, and Departement of Electrical and Computer Engineering, Duke University, Durham NC 27708, U.S.}

\date{\today}

\begin{abstract}
Classical nonlocality in conducting nanostructures has been shown to dramatically alter the linear optical response, by placing a fundamental limit on the maximum field enhancement that can be achieved. This limit directly extends to all nonlinear processes, which depend on field amplitudes.
A study of third-harmonic generation in metal film-coupled nanowires reveals that for sub-nanometer vacuum gaps the nonlocality enhances the effective nonlinearity by four orders of magnitude as the field penetrates deeper inside the metal than that predicted assuming a purely local electronic response.
\end{abstract}

\pacs{0000000}

\maketitle

Large optical nonlinearities are critical to photonic technologies. The exploitation of nonlinear processes at low power levels, and in highly integrated formats, requires materials with large nonlinear susceptibilities in configurations that offer efficient nonlinear conversion.
Metals have long been recognized as compelling candidates for nonlinear materials, as they possess nonlinear susceptibilities that are orders of magnitude larger than dielectric materials, and support surface plasmon modes that allow the light to become strongly confined and enhanced in deeply sub-wavelength volumes.
As a result, a major research effort has been targeted toward metal-dielectric composites\cite{Haus:1989fz,Iorsh:2012ei}, including metallo-dielectric stacks\cite{Bennink:1999wb,Larciprete,Katte:2011dd}, metamaterial composites\cite{Czaplicki:2013bd,Canfield:2004vi,Klein,Klein:2006tj,*Klein:2008tm,Ciraci:2012vw}, structured films\cite{Davoyan:2010jq} and surfaces\cite{Renger:2010je,Renger:2011wc,Wurtz:2011jc}.
While the absorption inherent to metals is generally considered to be detrimental for linear applications, it is far less critical for nonlinear optical applications because conversion rates are expected to be smaller than a fraction of a percent.
Although metals possess large nonlinearities, their high reflectivity has hindered their adoption as nonlinear optical materials. 

Even though metals may generally be opaque and highly reflective, metal nanostructures with features much smaller than the incident wavelength can interact strongly with light. For example, recent work on nano-structured metallic absorbers has shown that reflectivity and other optical characteristics can be modified substantially.
Structured nanoparticles, such as crosses, disks, or rods, deposited on a dielectric spacer atop a metal substrate can introduce an effective magnetic response that can impedance-match the surface to the vacuum, thus minimizing reflections and maximizing absorption\cite{Liu:2010kw,Landy:2008gy}.

In most key optical applications of metals, the nanoscale gaps between coupled metallic nanostructures are critical, with smaller gaps increasing local field enhancement and confinement. As gap size decreases to sub-nanometer scales, two basic issues must be dealt with: i) the conventional local description of the electronic response of strongly coupled metal nanoparticles breaks down, as the the dielectric constant of the metal acquires a wave-vector dependence; and ii) induced quantum currents turn the vacuum into a conductor as a result of quantum tunneling. 
In linear systems, the effect of classical nonlocality in gold-based film-coupled nanoparticles can limit the achievable field enhancement\cite{Ciraci:2012fp}.
The influence of quantum mechanical effects, in which electron tunneling reduces field enhancement in the gap region, has been investigated both theoretically\cite{Zuloaga:2009gm,Teperik:2013dd, Haus:2013vf} and experimentally\cite{Savage:2012by,Scholl:2012ge}.
In the extreme coupling regime, all of the models of electron response that extend beyond the local response tend to reduce the expected field enhancements, suggesting that nonlinear processes may also be significantly affected. 

In this letter we consider a configuration where the inclusion of nonlocality can dramatically enhance the nonlinear response by several orders of magnitude.This occurs because  surface charges begin to permeate the volume immediately beneath the surface of the metal, , thus allowing the fields to access the large, intrinsic third order nonlinearity typical of metals. 

%--------------FIGURE-----------------------------------
\begin{figure}
        \centering

                \includegraphics[width=0.45\textwidth]{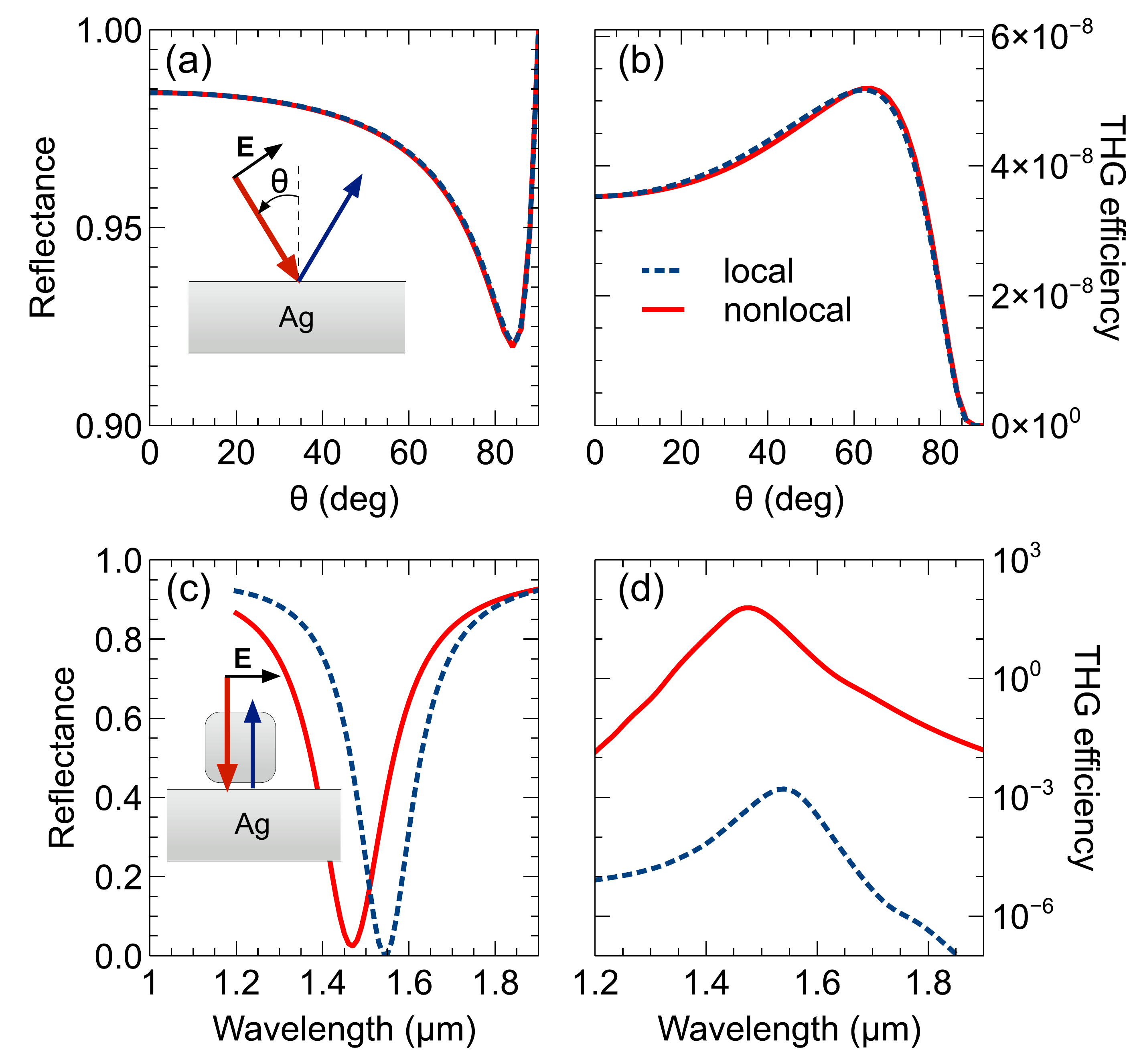}
       
        \caption{Comparison of THG efficiencies obtained using a local-response approximation and the hydrodynamic model for free electrons. (a) In the case of a bare silver film, the two model give the exact same result. (b) Film-coupled nanowires show a much higher THG efficiency when the hydrodynamic model is used.}
         \label{efficiency}
\end{figure}
%--------------------------------------------------

The process of third-harmonic generation (THG) can be described by means of the nonlinear wave equation.
For an electric field ${\bf E}$ oscillating at the angular frequency $\omega$ we have:
%-----------------EQUATION--------------------
\begin{equation}
\nabla  \times \nabla  \times {\bf{E}} - \frac{{{\omega ^2}}}{{{c^2}}}{\bf{E}} = {\omega ^2}{\mu _0}{\bf{P}},
\label{waveq}
\end{equation}
%--------------------------------------------------
where $c$ is the speed of light in free-space and $\mu_0$ is the magnetic permeably.
In Eq.~(\ref{waveq}), we explicitly write the polarization ${\bf P}$ of the medium as a source term.
The vector ${\bf P}$ may be subdivided into three contributions as follows: i) free electron response ${\bf P}_{\rm f}$, ii) bound electron response ${\bf P}_{\rm b}$ and iii) nonlinear response ${\bf P}_{\rm NL}$:
%-----------------EQUATION--------------------
 \begin{equation}
{\bf{P}} = {{\bf{P}}_{\rm{f}}} + {{\bf{P}}_{\rm{b}}} + {{\bf{P}}_{{\rm{NL}}}}.
\label{P}
 \end{equation}
%--------------------------------------------------

The free-electron portion is treated using the hydrodynamic model, an extension of the Drude model that accounts for the effect of electron pressure is taken into account\cite{Ciraci:2013dz}. In particular, the free-electron pressure gives rise to the nonlocal portion of the polarization, and may be determined from the equation:
%-----------------EQUATION--------------------
  \begin{equation}
- {\beta ^2}\nabla \left( {\nabla  \cdot {{\bf{P}}_{\rm{f}}}} \right) - \left( {{\omega ^2} + i\omega \gamma } \right){{\bf{P}}_{\rm{f}}} = \omega _{\rm{p}}^2{\varepsilon _0}{\bf{E}},
\label{free}
 \end{equation}
%-------------------------------------------------- 
where $\omega_{\rm p}$ and $\gamma$ are the plasma frequency and the collision rate, respectively, which also appear in the conventional Drude formula. The parameter $\beta$ is approximately the speed of sound in the Fermi-degenerate plasma of conduction electrons, that is $\beta^2=\frac{3}{5}v_F^2$.

The portion of the polarization due to the bound electrons is considered in the limit of the local-response approximation and it is described as a multipole Lorentz oscillator:
%-----------------EQUATION--------------------
  \begin{equation}
{{\bf{P}}_{\rm{b}}} = {\varepsilon _0}\left[ { - \sum\limits_j {\frac{{\omega _{{\rm{p}},j}^2}}{{{\omega ^2} - \omega _{0,j}^2 + i\omega {\gamma _j}}}} } \right]{\bf{E}},
\label{lorentz}
 \end{equation}
%-------------------------------------------------- 
 where $j$ is an index labeling the individual $d$-band to $sp$-band electron transitions occurring at $\omega_{0,j}$. 
The nonlinear contribution is a third-order function of the electric fields. 
For an isotropic material there is only one independent element of the susceptibility tensor  ${\chi _{ijkl}}(3\omega; \omega,\omega,\omega ) =\frac{1}{3}{\chi ^{(3)}}({\delta _{ij}}{\delta _{kl}} + {\delta _{ik}}{\delta _{jl}} + {\delta _{il}}{\delta _{kj}})$ so that the nonlinear polarization can be taken as\cite{Boyd:2006uq}:
%-----------------EQUATION--------------------
  \begin{equation}
{{\bf{P}}_{{\rm{NL}}}}  = {\varepsilon _0}{\chi ^{(3)}}\left( {{\bf{E}} \cdot {\bf{E}}} \right){\bf{E}}
\label{Pthg}
 \end{equation}
%-------------------------------------------------- 

We numerically solve the system of Eqs.~(\ref{waveq})-(\ref{Pthg}) in the undepleted pump approximation, using the finite-element method implemented in the commercially available software \textsc{Comsol} Multiphysics\cite{comsol}.
Since the equation for the free-electron polarization  is nonlocal, an additional set of boundary conditions are needed beyond the well-known tangential field continuity condition. 
The additional boundary conditions should be applied with some care as bound electrons are assumed to give a local contribution to the polarization vector. 
Following Ref.~\cite{Moreau:2013ei}, we impose that ${\bf P}_{\rm f}\cdot{\bf \hat{n}}=0$ at the surface, where ${\bf \hat{n}}$ is the unit vector normal to the surface.

%--------------FIGURE-----------------------------------
\begin{figure}
        \centering

                \includegraphics[width=0.4\textwidth]{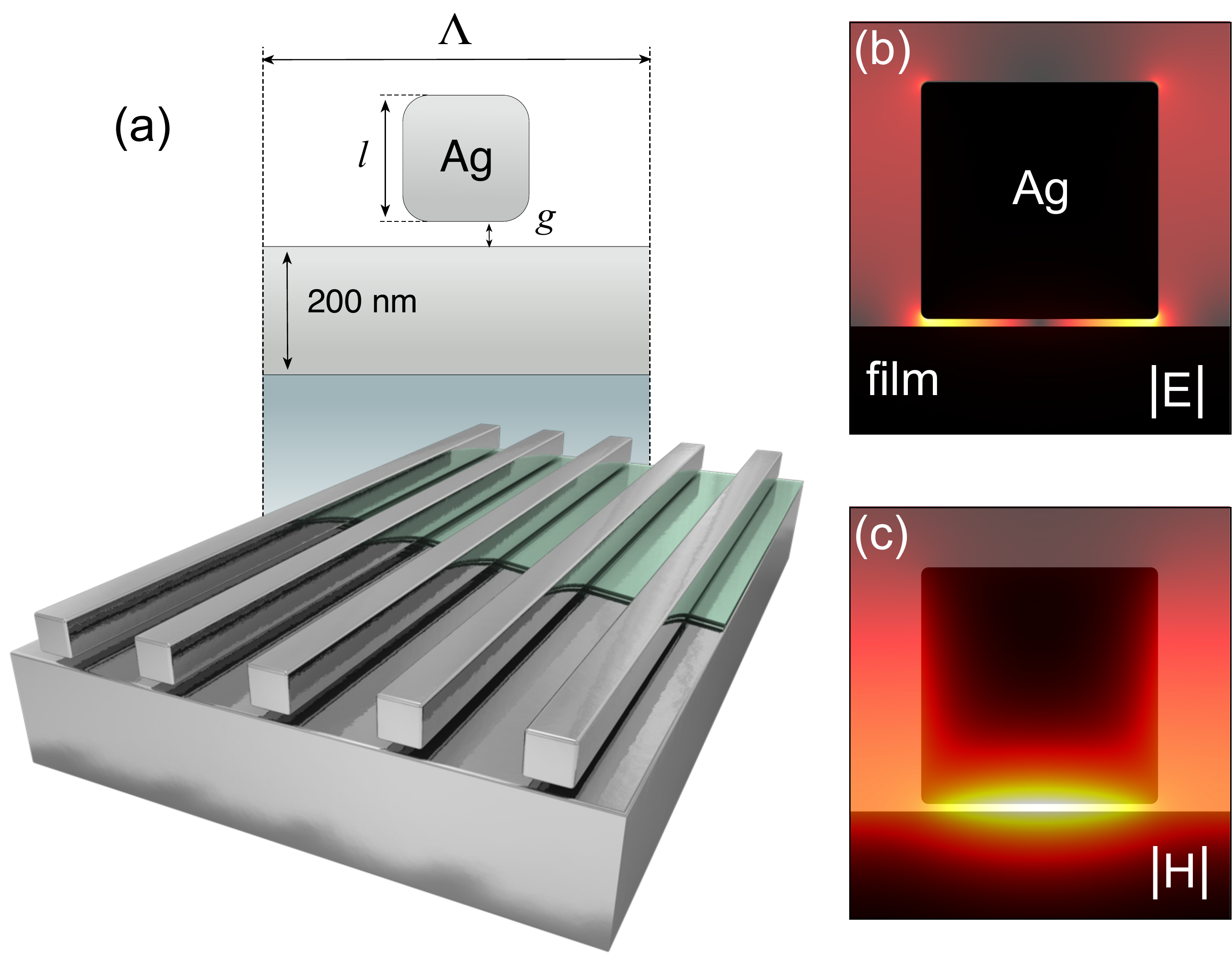}
       
        \caption{The film-coupled nanowire system. (a) Schematic of the geometry with array of nanowires. Cross-section of the magnetic field (b) and the electric field (c) from finite-element simulations, showing the fundamental cavity mode.}
         \label{geom}
\end{figure}
%--------------------------------------------------

We define the THG efficiency as:
%-----------------EQUATION--------------------
  \begin{equation}
\eta  = \frac{{I(3\omega )}}{{{I_0}(\omega ){{\left[ {\chi ^{(3)}}{E_0^2(\omega )} \right]}^2}}},
\label{eta}
 \end{equation}
%-------------------------------------------------- 
where $I_0(\omega)$ and $I(3\omega)$ are the intensity of the fundamental field and the measured intensity of the generated field respectively.
The THG efficiency does not depend either on the intensity of the fundamental wave, nor on the specific value of the bulk $\chi^{(3)}$ used.
That is, $\eta$ predicts the overall THG efficiency of the system, rather than just of a material with bulk $\chi^{(3)}$.
It is also useful to introduce the nonlocal enhancement factor as $\zeta=\eta_{HDM}/\eta_{LRA}$, where $\eta_{HDM}$ and $\eta_{LRA}$ are given by Eq.~(\ref{eta}) assuming a hydrodynamic model and local-response approximation, respectively.
From the previous definition we may define the effective nonlocal nonlinear susceptibility as:
%-----------------EQUATION--------------------
  \begin{equation}
\chi^{(3)}_{eff}= \zeta^\frac{1}{2} \chi^{(3)}.
\label{chieff}
 \end{equation}
%-------------------------------------------------- 

Having defined all of the equations and efficiencies, we consider the simple case of THG from a bare metal film.
To obtain a non-zero normal component with respect to the film surface, which allows the fields to access the nonlocal effects, we vary the incidence angle of the pumping field from 0$^\circ$ to 90$^\circ$.
Figure \ref{efficiency}a shows the THG efficiency for a silver film ($d=200$nm) in the local-response approximation ($\beta=0$), as well as for the case when the nonlocal response of the free electrons is taken into account through Eq.~(\ref{free}).
In the absence of nanostructured features the two curves overlap.

We next consider the geometry depicted in Fig.~\ref{geom}a, which consists of an array of infinitely long silver nanowires with square cross-sections of width $l$ and period $\Lambda$, separated by a distance $g$ from a silver film of thickness $d$.
For simplicity we assume the surrounding dielectric to be air. Each metal nanowire supports a strong resonance localized within the cavity formed between the bottom surface of the nanowire and the underlying metal film\cite{Moreau:2012uba}, as shown in Fig.~\ref{geom}b and Fig.~\ref{geom}c. 
The reflectivity of the patterned surface can reach values close to zero in correspondence with the peak resonance at normal incidence, as shown in Fig.~\ref{efficiency}c.
In this situation the electric field is squeezed into the gap between the film and the nanowire and locally enhanced in excess of two orders of magnitude with respect to the amplitude of the incident radiation.
Figure~\ref{efficiency}d shows the THG efficiency for the case of the structure depicted in Fig.~\ref{geom}, for a gap of $g=1$~nm.
The difference in THG between the local and nonlocal models is enormous. We predict that the by including the nonlocal response this geometry will foster an enhancement of THG of more than four orders of magnitude with respect to the local response.

%--------------FIGURE-----------------------------------
\begin{figure}
        \centering

                \includegraphics[width=0.45\textwidth]{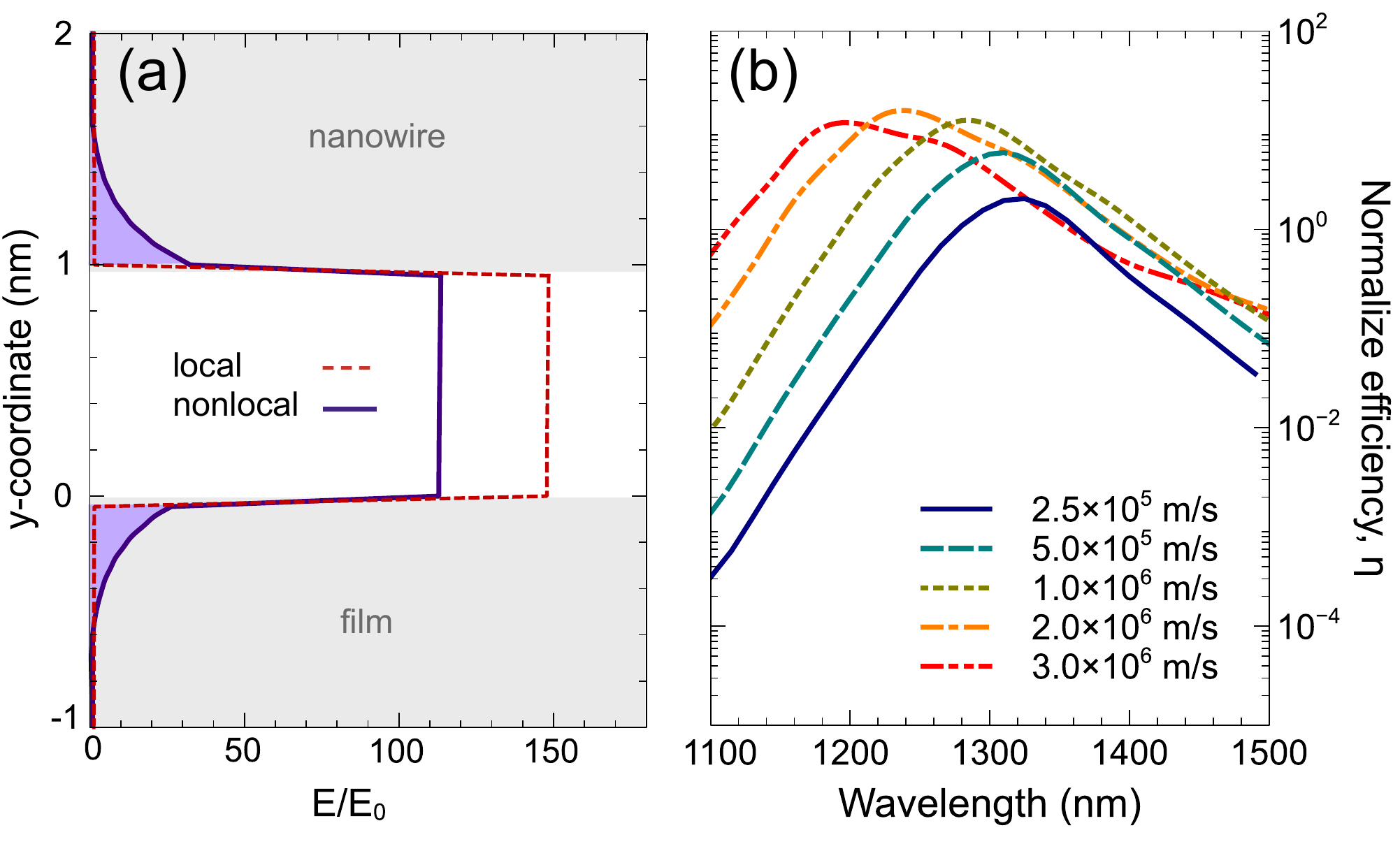}
       
        \caption{Effect of classical nonlocality. In (a) we plot the electric field amplitude on the cross-section along the gap for the local and nonlocal cases.
Outside the metal region the electric field enhancement is reduced when the nonlocal response is taken into account. (b) Nonlinear efficiency for different values of $\beta$.}
         \label{fields}
\end{figure}
%--------------------------------------------------

The origin of this striking contrast in THG efficiencies relates to behavior of the conduction electrons, which exhibit different screening properties at the sub-nanometer scale.
Figure~\ref{fields}a shows the electric field amplitude on the cross-section along the gap for the local and nonlocal cases.
Although outside the metal region the electric field enhancement is reduced when the nonlocal response is taken into account, the charges are smeared just beneath the metal surfaces, allowing the fields  to partially penetrate, reaching portions of the bound and core electrons responsible for the intrinsic third-order nonlinearity.
In Fig.~\ref{fields}b we plot the normalized efficiency for different values of the parameter $\beta$, ranging from $0.25$ to $3.0\times10^6$~m/s.
Depending on the precise value of the Fermi velocity, the efficiency can vary by more than one order of magnitude, but remains quite high.

Comparing these results to the case of a simple film suggests that the impact of the nonlocal response of free electrons may depend on the specific geometry and on the gap between the metal substrate and the nanowires.
That is, the nonlocal enhancement factor is a function of the gap, $\zeta =\zeta (g)$.

%--------------FIGURE-----------------------------------
\begin{figure}
        \centering

                \includegraphics[width=0.45\textwidth]{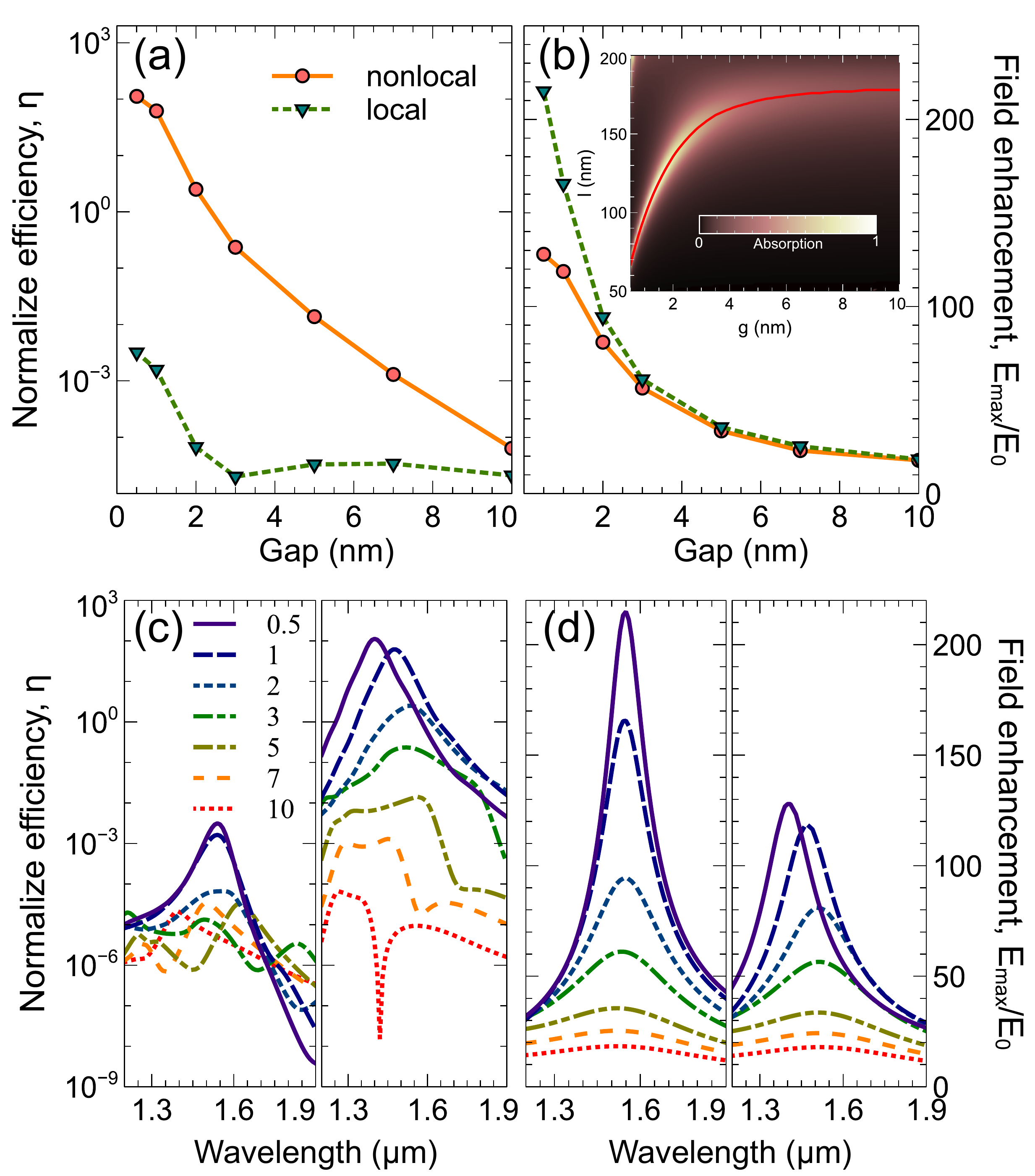}
       
        \caption{THG efficiencies (a) and field enhancement factors (b) are plotted as a function of the distance from the metal substrate. The different curves are obtained using the local-response approximation and the hydrodynamic model for free electrons, respectively. The inset shows the absorption map for the film-coupled nonowire system as a function of the parameters $l$ and $g$ for a constant wavelength $\lambda=1.55\mu$m and periodicity $\Lambda=200$nm. The field enhancement in the gap regions is plotted (white line) along the maximum absorption curve (red line). (c) and (d) show the full spectra of THG efficiencies and field enhancement factors.}
         \label{map}
\end{figure}
%--------------------------------------------------

In order to demonstrate this dependance, we perform calculations of THG as a function of the nanowire distance from the film.
However, one should be mindful of the fact that as $g$ increases the resonance blue-shifts, pushing the THG field into the UV range. 
In the inset of Fig.~\ref{map} we plot the linear absorption as a function of the parameters $l$ and $g$, keeping the wavelength of the pumping field constant and equal to $\lambda_{\rm FF}=1.55\mu$m.
The red curve represents the region where the absorption (reflectance) is a maximum (minimum) in the case of the local-response approximation.
The peak generation efficiencies $\eta_{LRA}$ and $\eta_{HDM}$ for both local and nonlocal cases respectively, are calculated along this line for several values of $g$ and plotted in Fig.~\ref{map}a along with the field enhancements in Fig.~\ref{map}b.
For large gaps, the efficiencies are of the same order of magnitude, meaning that the impact of the nonlocality is negligible.
However, the effect of the nonlocal, free-electron response becomes dominant as the gap closes, generating an enhancement of the THG process up to four orders of magnitude over that obtained from the local model.
By contrast, the field enhancement in the nonlocal case drops sensibly for smaller gaps.

In absolute terms, our system is capable of converting up to $0.0001\%$ of the incoming radiation for a peak pump power of $I_0(\omega)=50$~MW/cm$^2$ and a value of the nonlinear susceptibility for the silver of $\chi^{(3)}=10^{-18}$~m$^2$/V$^2$\cite{Boyd:2006uq}.
This value is nine orders of magnitude larger than THG conversion of a bare silver films, and several order of magnitude larger than THG conversion efficiencies reported in other metallic nano-composites\cite{Klein:2006tj,*Klein:2008tm}.

As the effective $\chi^{(3)}$ will depend on the distance of the nanowires and the film, this finding may be easily tested experimentally.
The same concept can be applied to the case of four-wave mixing.
The three-dimensional variant of the present structure--the film-coupled nanocubes--possess very similar resonances and can be easily fabricated by using colloidal methods\cite{Moreau:2012uba}.

Although the hydrodynamic model used to describe the nonlocal response of free-electrons may not be completely adequate to give an accurate quantum description of the subatomic realm, its simplicity enables qualitative and quantitative predictions for systems where a full-quantum approach is prohibitive. For example, in physical systems, the geometrical dimensions may exceed several hundreds of nanometers, making the computational domain for quantum calculations prohibitive. 
We have introduced an alternative way to experimentally investigate this model.
Our findings show a route to obtain efficient nonlinear processes that exceed other approaches by several orders of magnitude.

\begin{acknowledgments}
We thank D. de Ceglia, M. A. Vincenti, and J. W. Haus for helpful input and discussion. This work was supported by the Air Force Office of Scientific Research (AFOSR, Grant No. FA9550-09-1-0562) and by the Army Research Office through a Multidisciplinary University Research Initiative (Grant No. W911NF-09-1-0539).
\end{acknowledgments}

\end{document}